\documentstyle[12pt,aps,epsfig,prl]{revtex}

\relax  
\begin{document}
\preprint{prl} \draft

\title{Detailed single crystal EPR lineshape measurements for the single molecule magnets Fe$_8$Br and Mn$_{12}-$ac}

\bigskip

\author{S. Hill\cite{email}}
\address{Department of Physics, University of Florida, Gainesville, FL 32611}

\author{S. Maccagnano}
\address{Department of Physics, Montana State University, Bozeman, MT 59717}

\author{Kyungwha Park}
\address{School of Computational Science and Information Technology and Department of Chemistry, Florida State University, Tallahassee, FL 32606}

\author{R. M. Achey, J. M. North and N. S. Dalal}
\address{Department of Chemistry and National High Magnetic Field Laboratory,
Florida State University, Tallahassee, FL 32306}

\date{\today}
\maketitle

\smallskip

\begin{abstract}

\noindent{It is shown that our multi-high-frequency ($40-200$ GHz)
resonant cavity technique yields distortion-free high field EPR
spectra for single crystal samples of the uniaxial and biaxial
spin $S = 10$ single molecule magnets (SMMs)
[Mn$_{12}$O$_{12}$(CH$_3$COO)$_{16}$(H$_2$O)$_4$]$\cdot$2CH$_3$COOH$\cdot$4H$_2$O
and [Fe$_8$O$_2$(OH)$_{12}$(tacn)$_6$]Br$_8\cdot9$H$_2$O. The
observed lineshapes exhibit a pronounced dependence on
temperature, magnetic field, and the spin quantum numbers (M$_S$
values) associated with the levels involved in the transitions.
Measurements at many frequencies allow us to separate various
contributions to the EPR linewidths, including significant
$D-$strain, $g-$strain and broadening due to the random dipolar
fields of neighboring molecules. We also identify asymmetry in
some of the EPR lineshapes for Fe$_8$, and a previously unobserved
fine structure to some of the EPR lines for both the Fe$_8$ and
Mn$_{12}$ systems. These findings prove relevant to the mechanism
of quantum tunneling of magnetization in these SMMs. }

\end{abstract}

\smallskip


\clearpage
\centerline{\bf I. Introduction}
\bigskip

\noindent{Examples of "single molecule magnets" (SMMs) which have
attracted considerable recent interest include the uniaxial
[Mn$_{12}$O$_{12}$(CH$_3$COO)$_{16}$(H$_2$O)$_4$]$\cdot$2CH$_3$COOH$\cdot$4H$_2$O
(Mn$_{12}-$acetate, or Mn$_{12}-$ac) system, and the biaxial
[Fe$_8$O$_2$(OH)$_{12}$(tacn)$_6$]Br$_8\cdot9$H$_2$O (Fe$_8$Br)
system \cite{1,2,3,4,5,6,7,8,9,10}. Both of these SMMs exhibit
spin $S = 10$ ground states, with a dominant easy-axis type
anisotropy, $D\hat{S}_z^2$ (where {\em D} is negative $-$ see
section II), which provides a potential barrier (height
$\sim|D|S^2$) against the reversal of individual spins from
$M_S=+10$ to $-10$ (see Fig. 1). Recent studies have shown that
both systems possess an ability for their magnetic moments to
quantum tunnel through their respective anisotropy barriers (see
Fig. 1), and that this process dominates the magnetization
dynamics at low temperatures ({\em k}$_B$T$\ll|D|S^2$)
\cite{5,11,12,13,14,15,16,17,18}.}

Because of its significance as a fundamentally novel phenomenon,
as well as its potential for applications in nano-technology,
magnetic quantum tunneling (MQT) in SMMs has been extensively
studied lately (see {\em e.g.} \cite{5,11,12,13,14,15,16,17,18}),
yet its underlying mechanism is still not fully understood,
especially at a quantitative level. One possible reason is the
lack of precise information on the disposition of the individual
(electronic) spin energy levels. This is due to the fact that much
of the available data on the spin properties of these systems have
been obtained through DC and AC magnetization studies
\cite{14,15,16,Luis}, specific heat \cite{Fominaya}, neutron
scattering \cite{Hennion,Mirebeau,Caciuffo}, proton NMR
\cite{Lascialfari,Jang}, and muon spin relaxation
\cite{Lascialfari}, which cannot easily or directly probe the
properties of specific spin quantum levels.

An important breakthrough in the spectroscopy of SMMs has been the
development of a sensitive cavity perturbation technique for
carrying out multi-high-frequency ($40$ to $>200$ GHz) EPR
measurements on single crystals
\cite{hill1,jos2,hill2,saram,delbarco}. In particular, this
technique leads to at least a three orders of magnitude
improvement in detection sensitivity \cite{hill2} relative to
conventional single-pass multi-high-frequency EPR techniques
\cite{8,Caneschi,Barra2,Barra3,Barra4,Mukhin,Parks}, which opens
up many new avenues for probing SMMs. To date, our single crystal
investigations have focused primarily on the energy dependence of
the positions in magnetic field of EPR absorptions corresponding
to transitions between different spin quantum levels
\cite{hill1,jos2,hill2,saram}. In this article, we turn our
attention to the magnetic field, frequency, temperature and spin
quantum level (M$_S$) dependence of the EPR linewidths and, in
particular, the line shapes obtained from measurements on Fe$_8$Br
and Mn$_{12}-$ac. We note that a brief account has been presented
elsewhere \cite{Kyungwha}. However, this earlier report emphasizes
the theoretical foundations of the linewidth analysis, and does
not consider lineshapes.

Documented lack of resolution, sensitivity and/or various
experimental artifacts have been known to affect most high
frequency EPR investigations of Fe$_8$Br and Mn$_{12}-$ac to date.
It is commonly stated that observed EPR lineshapes cannot be
analyzed because they contain uncontrollable distortions due to
instrumental artifacts resulting from standing waves in the sample
probe, which cause a mixing of the dissipative and reactive sample
response \cite{Barra2,Barra3,Blinc}. This is most pronounced in
the case of the high magnetic field studies of Barra {\em et al.}
\cite{8,Barra2,Barra3,Barra4}; asymmetric lineshapes are generally
observed, making it extremely difficult to evaluate the true
resonance positions. Indeed, even for aligned polycrystals,
significant discrepancies can be noted between the observed and
calculated EPR transitions, with respect to both the peak
positions and intensities, as well as the line shapes and widths
\cite{Barra3}. As an example, it has been reported in
\cite{Barra3} that the linewidths in Fe$_8$Br approximately scale
as $M_S^2$, whereas our earlier studies \cite{saram,Kyungwha} have
shown a dominant linear dependence of the linewidths on $M_S$,
both for Fe$_8$Br and Mn$_{12}-$ac. This fact, along with the
asymmetric lineshapes observed in high field EPR studies, may
account for differences in the spin Hamiltonian parameters deduced
by single-pass ({\em i.e.} cavity-less) high-field EPR, as
compared to other techniques \cite{16,hill1,hill2,saram}. The zero
field coherent time-domain THz technique employed by Parks {\em et
al.} \cite{Parks}, is limited in energy resolution (resolution is
a fifth of the broadest EPR line width), as well as sensitivity.
Consequently, it is necessary to study a large sample made up from
unaligned microcrystals, and it is only possible to observe a few
transitions. Furthermore, experimental complications only permit a
linewidth analysis, {\em i.e.} no lineshape analysis is possible.
To date, the most precise SMM EPR lineshape analyses have been
reported by Mukhin {\em et al.}, for both Fe$_8$Br and
Mn$_{12}-$ac \cite{Mukhin}, using a zero-field coherent source
technique. Gaussian EPR lines are observed, which are attributed
to random dipolar fields in the sample. However, these studies are
also limited in sensitivity. Consequently, only a few transitions
are observed, even for very large polycrystalline samples.
Finally, conventional X-band EPR is limited by the fact that only
a few transitions close to the top of the barrier are accessible.
Furthermore, little or no information concerning the spin-spin
(T$_2$) and spin-lattice (T$_1$) relaxation times has been
possible in X-band \cite{Blinc}.

As we will show in this article, measurements on single crystals
are essential, as are several other unique aspects of our
technique. For example, contrary to recent assertions by Blinc
{\em et al.} \cite{Blinc}, we are able to obviate essentially all
instrumental artifacts that could influence/distort EPR line
shapes. In addition, our ability to measure at many frequencies,
temperatures and orientations enables us to distinguish between
several contributions to the EPR line widths and shapes. These
include: inter-SMM and hyperfine dipolar fields; distributions in
crystal field parameters; and spin-lattice interactions. Indeed,
each of these effects have been discussed in MQT literature, {\em
e.g.} dipolar fields \cite{dipolar}, distributed crystal field
parameters caused by dislocations \cite{Chudnovsky}, thermally
assisted tunneling \cite{14,Garanin1,Loss1}, the Landau-Zener
effect \cite{Loss2,Raedt,Dobrovitsky}, and hyperfine interactions
\cite{Prokofev,Wernsdorfer2}. Based on our recent theoretical work
\cite{Kyungwha}, we find that a distribution in the uniaxial
crystal field parameter, so called $D-$strain, contributes
significantly to the EPR linewidths for the Fe$_8$Br and
Mn$_{12}-ac$ systems. In addition, we observe an appreciable
inter-SMM dipolar broadening for Fe$_8$Br, and $g-$strain in
Mn$_{12}-ac$. The current work focuses on the EPR lineshapes. This
is particularly important in view of recent theoretical work
\cite{Chudnovsky} which considers local distortions in the SMM
site symmetries, and their role in the MQT phenomenon.

The article is organized as follows: in the next section (II), we
discuss the $S=10$ effective spin Hamiltonian in the context of
high-frequency EPR and the MQT phenomenon; in section III we
describe our experimental procedure; in section IV, we separately
present results for Fe$_8$Br (IV. A) and Mn$_{12}-$ac (IV. B); in
section V, we discuss the significance of our results, and we
summarize and conclude our findings in section VI.

\bigskip

\centerline{\bf II. The effective spin Hamiltonian}

\bigskip

\noindent{The effective spin Hamiltonian for the Mn$_{12}-ac$ and
Fe$_8$Br systems has the form:}

\bigskip

\centerline{\hfill\hfill $\hat{H}$ = $D\hat{S}_z^2$ +
$E(\hat{S}_x^2 - \hat{S}_y^2)$ +
$\mu_B$$\vec{B}$$\cdot$$\vec{\vec{g}}$$\cdot$$\vec{S}$ +
$\hat{O}_4$ + $\hat{H}'$ ; \hfill (1) \hfill}

\bigskip

\noindent{where $\vec{S}$ is the vector spin operator and
$\hat{S}_z$, $\hat{S}_x$ and $\hat{S}_y$ are the spin projection
operators along the easy, hard and intermediate axes respectively;
{\em D} is the uniaxial spin-spin coupling parameter and the
rhombic {\em E} ($<D/3$) term characterizes the magnetic
anisotropy in the plane perpendicular to the easy axis;
$\vec{\vec{g}}$ is the Land$\acute{e}$ g-tensor and $\vec{B}$ is
the applied magnetic field vector; finally, $\hat{O}_4$ denotes
weaker higher order terms of order four in the spin operators, and
$\hat{H}'$ represents additional perturbations such as those which
lead to EPR line broadening, as discussed in the introduction.
{\em D} is negative for both Mn$_{12}-ac$ ($=-0.457$ cm$^{-1}$
\cite{Mirebeau}) and Fe$_8$Br ($=-0.203$ cm$^{-1}$
\cite{Caciuffo}). The rhombic term $E$ is zero for Mn$_{12}-$ac,
due to its strictly axial ($S_4$) site symmetry, while $E=0.032$
cm$^{-1}$ for Fe$_8$Br \cite{Caciuffo}. For a cylindrically
symmetric system, for which $D\hat{S}_z^2$ would be the only
non-zero term in Eq. (1) in zero applied field, the energy
eigenstates may be labeled by the quantum number $M_S$ ($-S < M_S
< S$), which represents the projection of $S$ onto the easy axis;
the energy eigenvalues are then given by the expression $\epsilon
= DM_S^2$. This results in the energy barrier separating doubly
degenerate ($M_S=\pm i$, {\em i} = integer) spin "up" and "down"
states (see Fig. 1).}

In order for quantum tunneling to be possible, there must exist
finite terms in Eq. (1) that break the cylindrical symmetry ({\em
i.e.} do not commute with $\hat{S}_z$) and mix states in either
potential well, thereby lifting the $M_S=\pm i$ degeneracies. Such
terms include: the rhombic $E(\hat{S}_x^2 - \hat{S}_y^2)$ term;
higher order single-ion anisotropy terms in the transverse spin
operators ({\em e.g.} $\hat{S}_x^4$ and $\hat{S}_y^4$)
\cite{8,16,Mirebeau}; a transverse externally applied magnetic
field; transverse internal fields due to neighboring SMMs or
nuclei (contained in $\hat{H}'$ \cite{dipolar,Prokofev}); or local
distortions in the crystal field symmetry caused imperfections in
the crystal (also contained in $\hat{H}'$ \cite{Chudnovsky}) $-$
these may include dislocations, or a partial loss or disorder
among the ligand molecules. Pure quantum tunneling is observed
between the unperturbed $M_S=\pm 10$ degenerate ground states in
Fe$_8$Br below about 350 mK (see Fig. 1), and the observed
relaxation has been explained in terms of the $E(\hat{S}_x^2 -
\hat{S}_y^2)$ term in Eq. (1) \cite{Wernsdorfer3}. However, one
has to additionally consider the respective roles of inter-SMM
dipolar fields and nuclear hyperfine fields [$\hat{H}'$ in Eq.
(1)] in order to fully account for the observed tunneling in
Fe$_8$Br \cite{18,Prokofev}. For the Mn$_{12}-$ac system, the
situation is even more complex: pure ground state quantum
tunneling has not been observed in zero applied magnetic field,
and there is little consensus as to the origin of the apparent
(weak) transverse anisotropy which causes thermally assisted
quantum tunneling among levels near the top of the barrier (see
Fig. 1). The leading theories include: fourth order single ion
transverse anisotropy terms such as $\hat{S}_x^4 + \hat{S}_y^4$,
which are allowed under the axial ($S_4$) site symmetry
\cite{8,16,Mirebeau}; and local distortions in the axial crystal
fields caused, {\em e.g.} by dislocations \cite{Chudnovsky}.

A tunneling matrix element (term in Eq. 1 that does not commute
with $\hat{S}_z$) gives rise to eigenstates that are constructed
from symmetric and antisymmetric combinations of unperturbed
($M_S=\pm i$) states on either side of the barrier, and to a
lifting of the degeneracy (tunnel splitting) between these
symmetric and antisymmetric states. Thus, the tunneling matrix
elements modify the energy eigenvalues and should, therefore, be
detectable in an EPR experiment. However, for the systems of
interest, these effects are small. Consequently, great precision
is required. Although inelastic neutron scattering (INS)
measurements are capable of providing similar information, EPR
offers many advantages in terms of resolution, sensitivity (amount
of sample required) and, in particular, the fact that one can
systematically control the level mixing by means of an externally
applied magnetic field. In addition, the higher sensitivity of the
EPR technique assures that all of this may be achieved for a
single crystal. Several EPR (and INS) experiments have indicated
the presence of significant fourth order single-ion transverse
anisotropies for both Mn$_{12}-ac$ \cite{16,hill1,Barra2,Mukhin}
and Fe$_8$Br \cite{Barra3,Mukhin}. Nevertheless, fourth order
anisotropy {\em cannot} fully explain the observed spacing in
magnetic field of magnetization steps observed in low temperature
hysteresis experiments \cite{Friedman2}, and there remains
considerable disagreement as to the magnitude of the fourth order
terms \cite{16,hill1,Barra2,Barra3,Barra4,Mukhin,Bao}.
Consequently, higher resolution EPR measurements are necessary in
order to refine estimates of the Hamiltonian parameters, and to
determine whether EPR lineshapes contain additional information
concerning the MQT phenomenon.

\bigskip

\centerline{\bf III. Experimental}

\bigskip

\noindent{The $(2S + 1)-$fold quantum energy level structure
associated with a large molecular spin $S$ necessitates
spectroscopies spanning a wide frequency range. Furthermore, large
zero-field level splittings, due to the significant crystalline
anisotropy (large $|D|$) and large total spin {\em S}, demand the
use of frequencies and magnetic fields considerably higher (50 GHz
to 1 THz, and up to 10 tesla respectively) than those typically
used by the majority of EPR spectroscopists. The high degree of
sensitivity required for single crystal measurements is achieved
using a resonant cavity perturbation technique in combination with
a broad-band Millimeter-wave Vector Network Analyzer (MVNA)
exhibiting an exceptionally good signal-to-noise ratio; a detailed
description of this instrumentation can be found in ref.
\cite{hill2,hill3}. The MVNA is a phase sensitive, fully sweepable
(8 to 350 GHz \cite{extensions}), superheterodyne source/detection
system. Several sample probes couple the network analyzer to a
range of high sensitivity cavities ($Q-$factors of up to 25,000)
situated within the bore of a 7 tesla superconducting magnet. The
MVNA/cavity combination has been shown to exhibit a sensitivity of
at least $10^9$ spins$\cdot$G$^{-1}\cdot$s$^{-1}$, which is
comparable with the best narrow-band EPR spectrometers. This,
coupled with newly acquired sources and a split-pair magnet, allow
single crystal measurements at any frequency in the range from 8
to 250 GHz, at temperatures down to 1.5 K ($\pm0.01$K), and for
any geometrical combination of DC and AC field orientations up to
7 T (up to 45 T at the National High Magnetic Field Laboratory).}

The use of a narrow band cavity offers many important advantages
over non-resonant methods (see {\em e.g.} refs
\cite{8,Caneschi,Barra2,Barra3,Barra4}). Careful consideration
concerning the coupling of radiation to and from the cavity (via
waveguide), combined with the ability to study very small samples,
eliminates problems associated with standing waves in the sample
probe \cite{hill3}. This, in turn, eliminates a mixing of the
dissipative and reactive responses of the sample under
investigation and, when combined with a vector detection scheme,
enables faithful extraction of the true EPR lineshapes (both the
real and imaginary components), free from instrumental artifacts.
One other notable feature of the superheterodyne scheme is its
detection rate of 34 MHz; thus, there is no need for field
modulation, unlike other high field EPR methods
\cite{8,Caneschi,Barra2,Barra3}. Consequently, any raw data
displayed in this paper constitutes pure absorption. Finally, the
use of a cavity enables positioning of a single crystal sample
into a well defined electromagnetic field environment, {\em i.e.}
the orientations of the DC and AC magnetic fields relative to the
sample's crystallographic axes is precisely known, and the DC
field is very homogeneous over the tiny volume ($\ll 1$mm$^3$) of
the sample. In this way, additional contributions to the EPR
lineshapes are avoided due to, {\em e.g.} field inhomgeneities
over the sample volume, or slight mis-alignments of crystallites,
as may be the case for measurements on aligned powders.

Single Mn$_{12}-$ac and Fe$_8$Br crystals were grown using
literature methods \cite{2,7}. All measurements were performed in
the standard EPR configuration with the AC excitation field
transverse to the DC field. The Fe$_8$Br crystal used for the
majority of the studies reported in the following section was a
thin rhombic (acute angle $\sim60^o$) platelet with edges of
length $\sim 0.7$ mm, and thickness $\sim 0.17$ mm. The
Mn$_{12}-$ac crystal was needle shaped, with approximate
dimensions $1\times0.15\times0.05$ mm$^3$. Orientation of
Mn$_{12}-$ac crystal was straightforward, due to the fact that its
easy axis is defined by the needle axis of the sample. Alignment
of the Fe$_8$Br crystal necessitated measurements at many angles
\cite{saram}. All of the data presented in this paper were
obtained with the DC magnetic field aligned approximately parallel
to the samples' easy axes (hard axis data are presented elsewhere
\cite{hill1,saram}). Finally, temperature control, in the range 2
to 50 K, was achieved using a variable flow cryostat situated
within the bore of the superconducting magnet.

\bigskip

\centerline{\bf IV. Results}

\noindent{\bf A. Fe$_8$Br}

\noindent{Figure 2 shows raw temperature dependent data obtained
for the Fe$_8$Br SMM at a frequency of 116.931 GHz; the field is
applied along the easy axis to within an accuracy of about $1^o$.
A series of more-or-less evenly spaced resonant absorptions are
observed which become successively weaker with increasing field. A
detailed analysis of the frequency dependence of the EPR line
positions is published elsewhere \cite{saram}. Here we focus
instead on the line widths and shapes. In this frequency and field
range, and for this orientation, the levels involved in the
transitions are well separated and do not cross. Thus, M$_S$ is a
good quantum number, and we have labeled the resonances in Fig. 2
according to the levels involved in the transitions.}

It is apparent from Fig. 2 that the observed transitions exhibit
dramatic variations in linewidth, both as a function of M$_S$,
field and temperature. In order to characterize this width
dependence, it is first necessary to fit the line shapes. Fig. 3
shows several attempts to fit various transitions with both
Lorentzian and Gaussian functions; all of the data were obtained
from a single field sweep at a frequency of 89.035 GHz and a
temperature of 10 K, with the field close to alignment with the
easy axis ($8^o\pm1^o$). At the lowest fields, where the
resonances are broad (comparable to spacings) and therefore
overlap, fits involving multiple Gaussian or Lorentzian functions
are necessary (Fig. 3). As can be seen from this figure, the
Gaussian fit is noticeably better than the Lorentzian one,
suggesting that some form of inhomogeneous broadening mechanism is
at play. Although this has been noted by other authors
\cite{Mukhin}, these earlier measurements were performed on
polycrystalline powders, which were pressed into large pellets.
Consequently, the microcrystals in the pellet will have been
subjected to varying stresses that could be responsible for the
broad Gaussian lines. Here, we demonstrate that this is an
intrinsic feature of the Fe$_8$ SMMs (this is also the case for
Mn$_{12}$ $-$ see following section).

Fig. 4 shows Lorentzian and Gaussian fits to transitions excited
from successively higher M$_S$ levels (smaller absolute value of
M$_S$) from the same field sweep as the data in Fig. 3. The first
thing to note is the dramatic line narrowing with decreasing
absolute value of M$_S$. Also apparent (inspite of the reduced
signal-to-noise) is a slight change in lineshape, from
approximately Gaussian to a more Lorentzian form, as the lines
become narrower. The significance of this finding will be
discussed further below.

Figure 5 shows a compilation of Gaussian widths, versus M$_S$ (the
level from which the transition was excited), measured at many
different frequencies; all of the data were obtained at a
temperature of 10 K, and with the field applied at an angle
$\theta=8^o\pm1^o$ away from the easy axis. The pronounced (almost
linear) increase in EPR linewidth with M$_S$ is due to $D-$strain
\cite{Kyungwha}. Since the dominant contribution to the spin
Hamiltonian (Eq. 1) is proportional to M$_S^2$, energy differences
({\em i.e.} EPR transition frequencies) scale as M$_S$; hence
$D-$strain produces a linear width ($\Delta$energy) dependence on
M$_S$, which projects onto a width in field ($\Delta${\bf B}) that
also scales linearly with M$_S$. The rounding close to M$_S = 0$
is due to a convolution of the intrinsic lifetime (spin-lattice)
broadening and the M$_S$ dependent ($D-$strain) contribution. The
former is expected to produce a Lorentzian lineshape, thus
accounting for the apparent crossover from Gaussian to Lorentzian
EPR lines, as observed for the sharper transitions in Fig. 4. The
slight narrowing of the EPR widths in Fig. 5 as a function of
increasing frequency, and the weak asymmetry about M$_S = -1/2$,
is due to the fact that higher frequency (higher M$_S$)
transitions are observed at higher magnetic fields, where the
inter-SMM dipolar broadening is weaker (see discussion below). For
Fe$_8$Br, no measurable $g-$strain is observed. A theoretical
analysis of these trends may be found in ref. \cite{Kyungwha}.

The temperature dependent contribution to the linewidths for
Fe$_8$Br (Fig. 6) is dominated by random/fluctuating dipolar
fields at each SMM site, due to the magnetic moments of the
neighboring SMMs. At 116 GHz, the M$_S = -10$ to $-9$ transition
occurs very close to zero field ($\sim0.1$ T). Under zero field
cooling, the spin-up (M$_S = -10$) and spin-down (M$_S = +10$)
states will be equally populated. Hence, the random distribution
of spin-up and spin-down molecules results in a maximal
contribution to the Gaussian dipolar EPR line broadening, as
evidenced by the extremely large temperature dependent
contribution to the M$_S = -10$ to $-9$ transition in Fig. 6 as
T$\rightarrow0$. Upon raising the temperature, higher levels in
the scheme ($|$M$_S|<10$) become populated and the individual
moments begin to fluctuate more rapidly, thus leading to a
suppression of the dipolar broadening. The same qualitative
picture may be applied to the other transitions (M$_S = -9$ to
$-8$ {\em etc.}), provided that $k_B$T exceeds the Zeeman
splitting of the ground state. For $k_B$T less than the ground
state Zeeman splitting, the system polarizes and the dipole
fluctuations are suppressed, leading to an EPR line narrowing, as
confirmed for all but the M$_S = -10$ to $-9$ transition in Fig.
6. The crossover occurs when the M$_S = +10$ state starts to
depopulate, {\em i.e.} approximately when $k_B$T $=gS\mu_B${\bf
B}. The inset to Fig. 6 shows the crossover temperature
(T$_{CR}$), where the slope of the data in the main part of the
figure changes sign, versus the average magnetic field strength at
which each particular transition is observed (see Fig. 2). It is
found that T$_{CR}$ indeed scales linearly with {\bf B}.
Furthermore, the slope of the fit to the data in the inset to Fig.
6 is $14.4\pm1$ K$\cdot$T$^{-1} = gS\mu_B/k_B$, corresponding very
well to $S=10$ and $g\approx2$ to within the experimental error,
thus supporting the picture of dipolar broadening. T$_{CR}$ for
the 116 GHz M$_S = -10$ to $-9$ transition in Fig. 6 is below the
minimum temperature of our current spectrometer (T$_{min}\sim 1.2$
K). Consequently, we were unable to verify whether this line
eventually narrows when the spin system polarizes. However, higher
frequency/field data, for which T$_{CR}$ should exceed the minimum
temperature of our spectrometer, exhibit a similar line broadening
for the M$_S = -10$ to $-9$ transition, at all temperatures
investigated (see Fig. 8 below). At present, the origin of this
effect is not fully understood.

In Fig. 7, we show the temperature dependent shifts of the EPR
line positions obtained from Fig. 2. When compared to Fig. 6, it
is noticeable that there is a similar crossover in the temperature
dependence for transitions from levels with $|$M$_S|<10$.
Therefore, this low-temperature limiting behavior may also be
attributed to the polarization of the spin system. The sign of the
shift indicates that the field at each SMM site, due to
neighboring SMMs, opposes the applied field, {\em i.e.} it is
opposite to the polarization direction. This is to be expected for
the Fe$_8$Br structure, in which the Fe$_8$ molecules reside on
the corners of a slightly distorted cube \cite{Hidalgo}. From the
magnitudes of the $|$M$_S|<10$ line shifts, we estimate dipolar
fields at each SMM site of a few hundred Gauss ($\sim 300$ G). The
origin of the continued shift of the M$_S = -10$ to $-9$
transition to lower magnetic fields is, at present, not well
understood.

A comprehensive theoretical analysis of this linewidth data has
been published elsewhere \cite{Kyungwha}. From such an analysis
for the sample used in obtaining the data in Fig. 2, we estimate a
distribution in {\em D} of $\sigma_D=0.01D$, negligible
$g-$strain, and a dipolar contribution to the broadening which
assumes a nearest neighbor SMM separation of 12 $\dot{A}$
\cite{simplify}, which is in agreement with the accepted range of
near neighbor distances of between $10.5-14 \dot{A}$. Within
acceptable ranges, no other combination of parameter values is
able to produce the same quality of fit simultaneously to all of
the frequencies and temperatures studied for this particular
sample, even though other parameter values may produce good fits
for a particular measurement. This emphasizes the power of our
multi-frequency single crystal EPR technique.

Very recent studies on a different sample from the one used in the
measurements described above (similar shape and size, and obtained
from the same synthesis \cite{age}), reveal unusual lineshape
characteristics, particularly at low temperatures and for
transitions involving levels close to the bottom of the potential
wells ({\em e.g.} M$_S=-10 \rightarrow -9$ and M$_S=-9 \rightarrow
-8 $,..). This is illustrated in Fig. 8, for data obtained at a
frequency of 145.937 GHz, with the field aligned precisely along
the easy axis. At the somewhat elevated temperature of 10 K (Fig.
8a), the M$_S=-10 \rightarrow -9$ transition shows appreciable
asymmetry: this resonance has a Lorentzian (dotted curve) tail on
the low field side, and closer to a Gaussian (dashed curve) tail
on the high field side; furthermore, the data lie somewhat to the
right of the fits on the high field side. Subsequent resonances
(M$_S=-9 \rightarrow -8$, M$_S=-8 \rightarrow -7$, {\em etc.})
appear to be less asymmetric, and the Gaussian fit improves.
Recent theoretical studies have considered, in some detail, the
effects of different types of defect structure on EPR lineshapes,
and how these defects may influence the MQT phenomenon in SMMs
(see section V) \cite{Chudnovsky,Kyungwha2}. Both Lorentzian and
Gaussian broadening are predicted \cite{Chudnovsky}, as well as
asymmetry \cite{Kyungwha2} (see discussion in section V).

Examination of the M$_S=-10 \rightarrow -9$ transition at lower
temperatures (Fig. 8b) reveals a fine structure to the line, which
is strongly temperature dependent, i.e. the curves are not smooth,
but exhibit a series of distinct distortions. We suspect that this
fine structure and the asymmetry is connected with crystal
defects, since it is not observed for all samples \cite{age}. A
direct comparison between the data in Fig. 8 and data for the
first sample (Figs 3 and 4) is complicated by the fact that the
linewidths are substantially broader for this second sample
(increased $D-$strain). We note that many SMM compounds are known
to lose solvent molecules \cite{age,Christou}, which would lead to
a degredation of sample quality over time and may, in fact,
provide a natural explanation for the $D-$strain. Nevertheless,
the M$_S=-10 \rightarrow -9$ transition in Fig. 2 (lowest field
resonance) also exhibits fine structure at the lowest
temperatures, and fine structure is also apparent in data
published by Barra {\em et al}. \cite{Barra3}. It is possible that
the fine structure is indicative of a finite number of SMM site
symmetries, as opposed to a smooth distribution. The dislocation
theory of Garanin and Chudnovsky \cite{Chudnovsky} leads to long
range strains in the crystal, thus giving rise to a smooth
distribution in site symmetries. Therefore, the fine structure we
observe may signify a finite number of configurations involving
Fe$_8$ molecules coordinated by different numbers or arrangements
of ligand molecules. It should be stressed that the data presented
in Fig. 8 truly represent the intrinsic EPR lineshapes, {\em i.e.}
any instrumental effects would be systematic and, therefore,
influence all of the lineshapes equivalently. Furthermore, the
noise level for these measurements is well below the resolution of
the figure. Clearly, additional studies on a range of samples
prepared in different ways, {\em i.e.} different sample qualities,
will be required to fully resolve these issues $-$ such studies
are currently in progress.

\bigskip

\noindent{\bf B. Mn$_{12}-$ac}

\bigskip

\noindent{Figure 9 shows raw data obtained for Mn$_{12}-$ac at a
frequency of 189.123 GHz, and at temperatures between 10K and 35K;
the field is applied parallel to the easy axis to within $0.2^o$.
The first point to note is the diminished signal-to-noise for this
data, when compared to Fig. 2. There are several reasons for this:
first, the dynamic range of our spectrometer is reduced at this
higher frequency of 189 GHz; second, the volume of the sample used
in this investigation was somewhat smaller than that of the
Fe$_8$Br sample used for the studies in the previous section;
finally, due to the increased energy scale of the Mn$_{12}-$ac
system (larger $|D|$), the observed transitions correspond to
levels considerably higher in energy above the ground state than
would be the case for Fe$_8$Br at comparable frequencies and
fields. Therefore, this constitutes a much more challenging
experiment. Nevertheless, clear resonances are observed, which
also have Gaussian lineshapes (solid curves), and once again
exhibit a pronounced linewidth dependence on M$_S$ (the level from
which the transition was excited). For comparison, the inset to
Fig. 9 shows attempts to fit the first two transitions with
Gaussian and Lorentzian functions; the Gaussian fit is noticeably
better.}

Figure 10 plots the M$_S$ dependence of the Gaussian EPR
linewidths for measurements made at several different frequencies.
The linear M$_S$ contribution to the widths once again indicates a
significant distribution in {\em D} ($\sigma_D \sim 0.02D$).
However, the dipolar contribution to the linewidths is
considerably weaker for Mn$_{12}-$ac than for Fe$_8$Br. This is
initially apparent from the weak frequency dependence observed
amongst the widths for transitions with $|$M$_S|<5$, and the fact
that the frequency dependence observed for higher $|$M$_S|$
transitions is opposite to what was observed for Fe$_8$ in Fig. 5,
{\em i.e.} the higher frequency transitions are broader. The
origin of this effect has to do with g-strain, and will be
discussed further below.

In Fig. 11, we plot the temperature dependence of the linewidths
observed at a frequency of 189.123 GHz. A similar trend can be
observed for the low $|$M$_S|$ transitions for both Mn$_{12}-$ac
and Fe$_8$Br (decreasing width with decreasing temperature).
However, the pronounced line narrowing observed for the higher
$|$M$_S|$ transitions at lower temperatures, in Fe$_8$Br, is
absent in Mn$_{12}-$ac. This provides further evidence that the
dipolar contribution to the linewidths is considerably weaker in
Mn$_{12}-$ac. In contrast, the M$_S$ dependence of the widths is
stronger for Mn$_{12}-$ac, due to the increased $D-$strain
($\sigma_D\approx0.02D$, as opposed to $\sigma_D\approx0.01D$ for
Fe$_8$Br). The weaker dipolar broadening is consistent with an
increased inter-SMM separation for Mn$_{12}-$ac \cite{2}.

The pronounced frequency dependence of the larger $|$M$_S|$ ($>5$)
transitions, as seen in Fig. 10, is attributable to $g-$strain.
Each transition shifts to higher field with increasing frequency,
thereby amplifying the effect of a distribution in g $-$ hence the
increased linewidths with increasing frequency. Additional support
for $g-$strain comes from the measured anisotropy in the
$g$-tensor, both from our previous work \cite{hill1} and that of
others \cite{Caneschi,Barra2}. Using accepted values for the
inter-SMM separations, it is possible to simulate all data
obtained for a particular sample with a single pair of (sample
dependent) $D-$ and $g-$strain parameters; $\sigma_D=0.02D$ and
$\sigma_g=0.008g$ are typical for many of the Mn$_{12}-$ac samples
we have studied \cite{Kyungwha}.

Finally, in Fig. 12, we show magnifications of several of the
resonances from Fig. 9. Once again, it is apparent that the EPR
lines exhibit a fine structure. In particular, the M$_S = -6$ to
$-5$ and M$_S = -5$ to $-4$ transitions show a clear double peak
structure; the best attempts at single Gaussian fits (dotted
curves) are included in the figure for comparison. Similar
behavior was observed at other frequencies. Once again, these
findings hint at the possibility that there may exist a finite
number of configurations involving Mn$_{12}-$ac molecules
coordinated by different numbers or arrangements of ligand
molecules. We note that the presence of various isomeric forms of
Mn$_{12}$, whereby the arrangements of the ligands differ
slightly, is well established both in the acetate \cite{Werns} and
in Mn$_{12}$ SMMs involving other ligands \cite{Christou2}.

\bigskip

\centerline{\bf V. Discussion}

\noindent{The unambiguous demonstration of $D-$strain in Fe$_8$Br
and Mn$_{12}-$ac indicates the presence of significant sample
inhomogeneities or imperfections, {\em i.e.} the molecules are
subjected to varying anisotropy. A natural explanation for this
could be a loss of solvent molecules, or disorder among the
ligands; indeed a slight degredation of sample quality over time
has been noted, and it is known that the crystals do lose solvent
\cite{Christou}. Another possibility has recently been proposed by
Garanin and Chudnovsky \cite{Chudnovsky}, involving dislocations
in the crystals which act to locally distort the site symmetries
of individual SMMs. What is more, Garanin and Chudnovsky have
claimed that the QTM phenomenon in SMMs finds a natural
explanation in terms of these dislocations. As mentioned in the
introduction, QTM requires symmetry breaking terms in the
Hamiltonian (Eq. 1). Local rotations of the magnetic anisotropy
axes due to dislocations have two effects: they generate
transverse second order crystal field terms, of the form
$E(\hat{S}_x^2 - \hat{S}_y^2)$, which facilitate QTM between M$_S$
states differing by an even integer; and they result in an
effective local transverse magnetic field (under application of
field$\|z$) that unfreezes odd tunneling resonances, thereby
solving the conundrum as to why both even and odd tunneling
resonances are always observed in Fe$_8$Br and Mn$_{12}-$ac. It
has also been shown that long-range strains produced by
dislocations result in broad distributions of relaxation times,
something which is also observed experimentally in Mn$_{12}-$ac
\cite{Chudnovsky}. Indeed, assuming realistic concentrations of
dislocations, Garanin and Chudnovsky \cite{Chudnovsky} have been
able to account for the experimentally observed tunneling in
Mn$_{12}-$ac \cite{Friedman2}, and predict a distribution in the
uniaxial spin-spin coupling parameter $\sigma_D\approx0.027D$,
which is of the same order as we have found from linewidth
analyses ($\sigma_D\approx0.01-0.02D$). It should be pointed out,
however, that any disorder amongst the ligand molecules, as well
as any loss of solvent molecules, may be expected to produce many
of the same effects as those proposed by Garanin and Chudnovsky
\cite{Chudnovsky}, {\em i.e.} disorder amongst the peripheral
molecules surrounding the SMMs will result in local variations in
site symmetry.

Careful theoretical consideration of different forms of
dislocations (edge, screw, correlated, uncorrelated, {\em etc}..)
indicate that the resultant $D-$strain can produce both Gaussian
and Lorentzian EPR lineshapes \cite{Chudnovsky}. Furthermore,
recent work by Park {\em et al.} \cite{Kyungwha2}, predicts
asymmetry in some of the EPR lineshapes; one may think of this as
arising due to the fact that a dislocation causes an asymmetric
distribution in $D$, {\em i.e.} it can only reduce $D$, not
enhance it. Clearly, certain features of the results presented
here, particularly in Fig. 8a, may confirm some of these
predictions. So far, our attempts to simulate data have been
limited to $D-$strain, $g-$strain and dipolar broadening. The $D-$
and $g-$strains should be correlated, although we have yet to
verify this. Simulated $E-$strain, as well as strain in the fourth
order anisotropy, did not affect our simulations to within
sensible ranges of these parameters. We have also ignored
hyperfine effects, since these are irrelevant in the case of the
Fe$_8$Br ($^{56}$Fe has no nuclear spin), and exchange narrowing
in Mn$_{12}-$ac is believed to diminish the hyperfine splitting
\cite{Kyungwha}. At present, most of our data have been obtained
for just a few samples of each of Fe$_8$Br and Mn$_{12}-$ac.
Further work is in progress on a range of different SMMs, with the
goal of resolving some of these issues, {\em e.g.} whether solvent
loss or ligand disorder is the main reason for the observed
$D-$strain, or whether EPR lineshapes correlate with various
theoretical works in progress dealing with dislocation mediated
QTM \cite{Kyungwha2}.

Clearly, if the QTM mechanism is mediated by disorder, then single
crystal EPR offers the most direct means of providing quantitative
information with which to test relevant theoretical predictions.
However, it is essential that EPR lineshapes obtained from
experiments are free from the experimental artifacts which have
been known to hamper high frequency EPR measurements. Indeed,
given the strong M$_S$ dependence of the EPR line shapes and
widths, one has to question recent attempts to determine the
crystal field parameters to fourth order from high frequency EPR.
Inspection of the fits to data for Fe$_8$Br and Mn$_{12}-$ac by
Barra {\em et al.} \cite{8,Barra2,Barra3,Barra4}, reveal
considerable discrepancies in the positions of low $|$M$_S|$
transitions. Because of the smaller linewidths associated with the
low $|$M$_S|$ transitions, one can be far more confident as to the
experimentally determined resonance field from such transitions,
than one can be for the considerably broader high $|$M$_S|$
resonances. Field modulation techniques compound this problem, as
do studies on aligned polycrystals which give rise to asymmetric
distributions in the orientations of the individual crystallites.
Indeed, our most recent investigations indicate that the high
frequency EPR technique described in this article (section III and
\cite{hill2,hill3}) yields crystal field parameters that are in
much closer agreement with inelastic neutron scattering
measurements than conventional high field EPR studies
\cite{hill4}.

\bigskip

\centerline{\bf VI. Summary and conclusions}

We have used a multi-high frequency cavity perturbation technique
to measure EPR lineshapes for oriented single crystal samples of
the Fe$_8$Br and Mn$_{12}-$ac SMMs. The use of a narrow band
cavity offers many important advantages over non-resonant methods:
careful consideration concerning the coupling of radiation to and
from the cavity (via waveguide), combined with the ability to
study very small samples, eliminates mixing of the dissipative and
reactive responses of the sample under investigation, and enables
faithful extraction of the true EPR lineshapes. Our ability to
measure at many frequencies, temperatures and orientations further
allows us to distinguish between several contributions to the EPR
line widths and shapes. These include: inter-SMM and hyperfine
dipolar fields; distributions in crystal field parameters; and
spin-lattice interactions. Each of these effects have been
discussed in the literature in terms playing a possible role in
the MQT phenomenon. In addition to significant $D-$strain, which
is indicative of a distribution in the SMM site symmetries, we
have detected a previously unobserved fine structure to some of
the resonances for both Fe$_8$Br and Mn$_{12}-$ac SMMs.

\bigskip

\bigskip

\centerline{\bf VII. Acknowledgements}

\bigskip

\noindent{We would like to thank George Christou, David
Hendrickson and Andrew Kent for useful discussion. This work was
supported by the National Science Foundation (DMR0103290 and
DMR0196430). S. H. is a Cottrell Scholar of the Research
Corporation.}


\clearpage

\noindent{{\bf Figure captions}}

\bigskip

Fig. 1 a) Schematic of the double potential well for a spin $S=10$
system in zero applied magnetic field, with a dominant easy axis
crystalline anisotropy $D\hat{S}_z^2$, with $D<0$; the M$_S=\pm10$
"spin-up" and "spin-down" states lie lowest in energy. Pure QTM,
directly between the unperturbed M$_S=\pm10$ states (as
indicated), is possible if there exist terms in the Hamiltonian
(Eq. 1) which mix these two levels. In the absence of pure QTM, a
thermally assisted form of QTM may occur if states higher up in
the scheme are mixed. b) Application of a magnetic field biases
the wells, favoring population of the $-$M$_S$ states. Whenever
levels in the two wells come into resonance, direct QTM is
possible, and accelerated relaxation occurs between the spin-up
and the spin-down states, {\em i.e.} the relaxation may be
considerably faster than the classical thermally activated
relaxation over the top of the barrier. Resonant QTM has been
observed in the form of sharp steps in the hysteresis loops of
SMMs \cite{16}.

\bigskip

Fig. 2. Temperature (indicated in the figure) dependent microwave
absorption amplitude for Fe$_8$Br, as a function of magnetic
field, for measurements at 116.931 GHz; the field is applied
parallel to the easy axis to within an accuracy of $1^o$. Sharp
resonances are observed, which correspond to EPR transitions
between the $2S+1$ spin states of the system $-$ these are
labelled in the figure.

\bigskip

Fig. 3. A close up of several Fe$_8$Br resonances obtained at 10 K
and a frequency of 89.035 GHz, with the field applied
approximately parallel to the easy axis; the levels involved in
the transitions are indicated in the figure. The solid and dashed
curves represent respective Gaussian and Lorentzian fits to the
data.

\bigskip

Fig. 4. Further Gaussian and Lorentzian fits to Fe$_8$Br data
(same experimental conditions as Fig. 3), illustrating the M$_S$
(level from which the transition was excited $-$ indicated in the
figure ) dependence of the widths. The horizontal scale has been
offset and normalized to the widths of each resonance $-$ the
actual widths are indicated in the figure.

\bigskip

Fig. 5. A compilation of the M$_S$ (level from which the
transition was excited) dependence of the Gaussian EPR linewidths
obtained for Fe$_8$Br at 10 K and many different frequencies
(indicated in the figure); see text for a discussion of the data.

\bigskip

Fig. 6. Temperature dependence of the Gaussian EPR linewidths
obtained for different spin transitions (indicated in the figure);
the data were obtained at a frequency of 116.931 GHz, with the
field applied parallel to the easy axis. The inset plots the field
dependence of the crossover temperature (T$_{CR}$) $-$ the
temperature at which the linewidths reach a maximum $-$ for the
M$_S=-9$ to $-8$, $-8$ to $-7$, $-7$ to $-6$ and $-6$ to $-5$
transitions.

\bigskip

Fig. 7. Temperature dependence of the EPR line positions obtained
for different spin transitions (indicated in the figure); the data
were obtained at a frequency of 116.931 GHz, with the field
applied parallel to the easy axis. The line shifts [$\mu_o$H(T)]
are plotted relative to the respective line positions at 50 K
[$\mu_o$H(50 K)].

\bigskip

Fig. 8a) A close up of several Fe$_8$Br resonances obtained at 10
K and a frequency of 145.937 GHz, with the field applied exactly
parallel to the easy axis; the levels involved in the transitions
are indicated in the figure. The solid and dashed curves represent
respective Gaussian and Lorentzian fits to the data. b)
Temperature dependence of the M$_S=-10$ to $-9$ transition in (a)
illustrating the fine structure to this line.

\bigskip

Fig. 9. Temperature (indicated in the figure) dependent microwave
absorption amplitude for Mn$_{12}-$ac, as a function of magnetic
field, for measurements at 189.123 GHz; the field is applied
parallel to the easy axis. Sharp resonances are observed, which
correspond to EPR transitions between the $2S+1$ spin states
(indicated in the figure) of the system. The solid curves are
Gaussian fits to the data; the inset shows attempts to fit both
Gaussian (solid curve) and Lorentzian (dashed curve) functions to
some of the data in the main part of the figure.

\bigskip

Fig. 10.  A compilation of the M$_S$ (level from which the
transition was excited) dependence of the Gaussian EPR linewidths
obtained for Mn$_{12}-$ac at 20 K and several different
frequencies (indicated in the figure); see text for a discussion
of the data.

\bigskip

Fig. 11. Temperature dependence of the Gaussian EPR linewidths
obtained for several spin transitions (indicated in the figure) in
Mn$_{12}-$ac; the data were obtained at a frequency of 189.123
GHz, with the field applied parallel to the easy axis. For
comparison, see data for Fe$_8$Br in Fig. 6.

\bigskip

Fig. 12. Magnifications of several of the 189.123 GHz traces shown
in Fig. 9 (transitions and temperatures are indicated in the
figure), illustrating a fine structure.

\clearpage

\begin{figure}
\centerline{\epsfig{figure=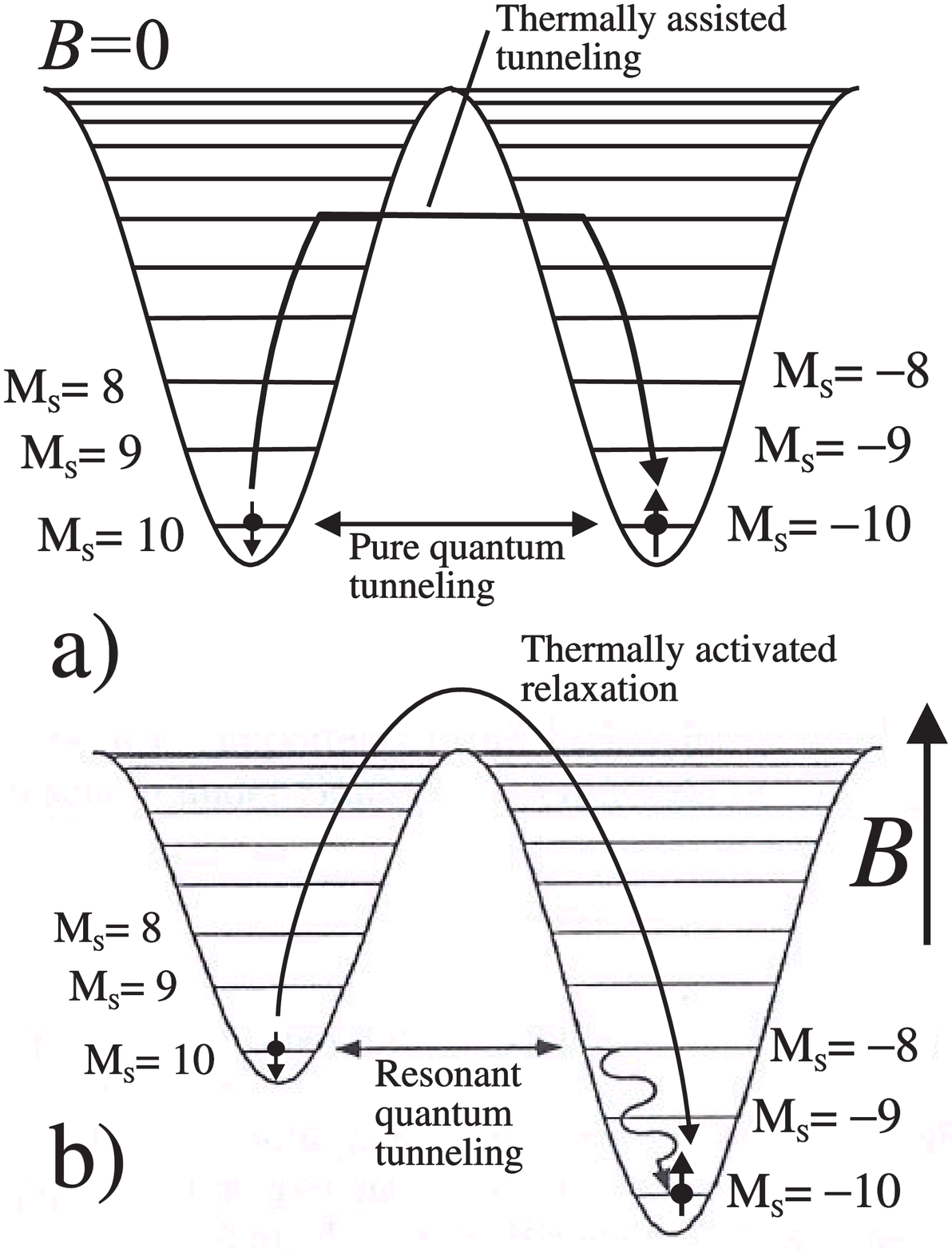,width=120mm}}
\bigskip
\caption{Hill {\em et al.}} \label{Fig. 1}
\end{figure}

\clearpage

\begin{figure}
\centerline{\epsfig{figure=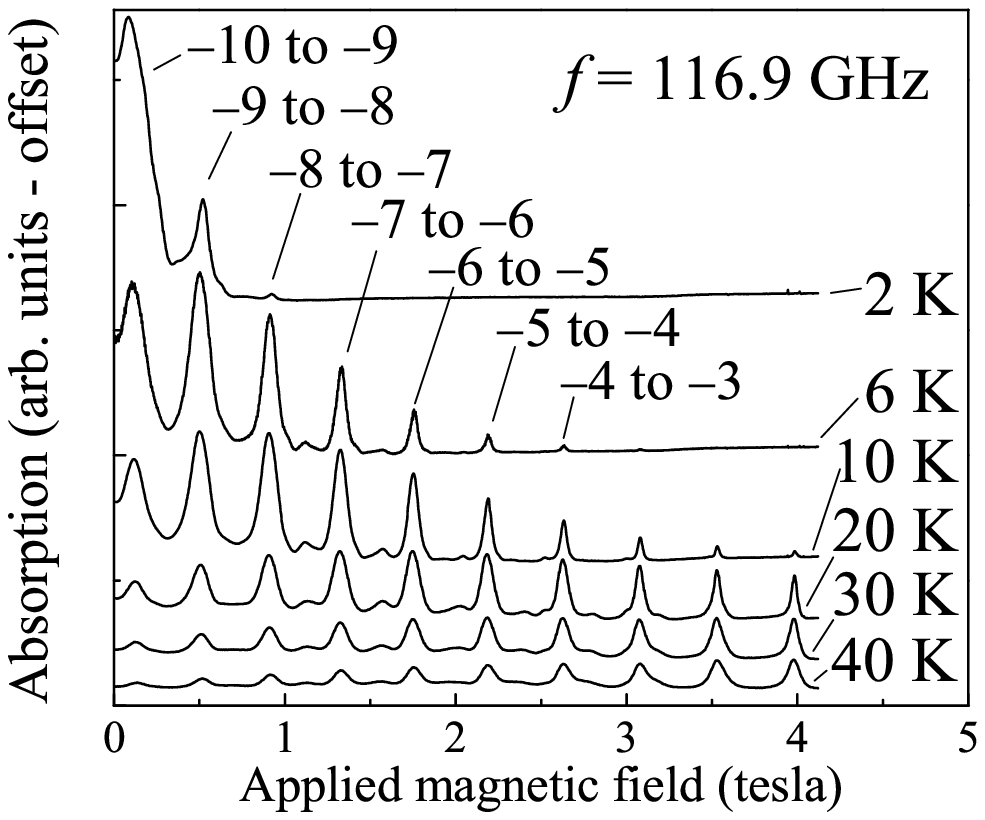,width=160mm}}
\bigskip
\caption{Hill {\em et al.}} \label{Fig. 2}
\end{figure}

\clearpage

\begin{figure}
\centerline{\epsfig{figure=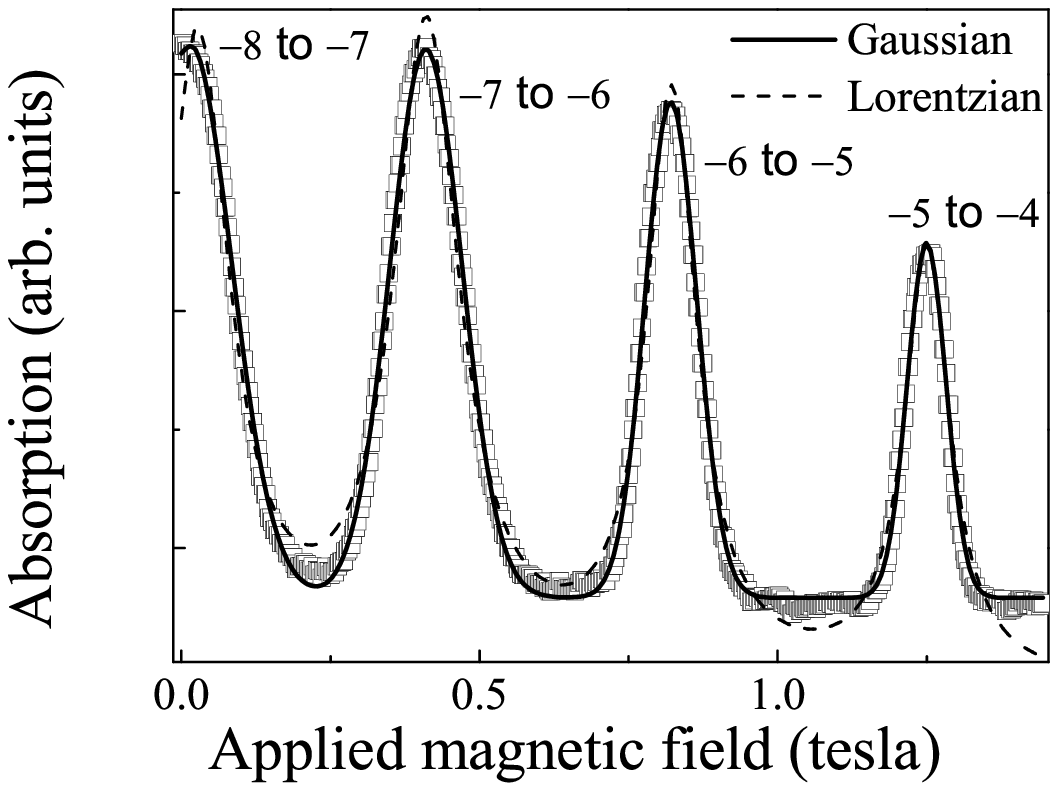,width=160mm}}
\bigskip
\caption{Hill {\em et al.}} \label{Fig. 3}
\end{figure}

\clearpage

\begin{figure}
\centerline{\epsfig{figure=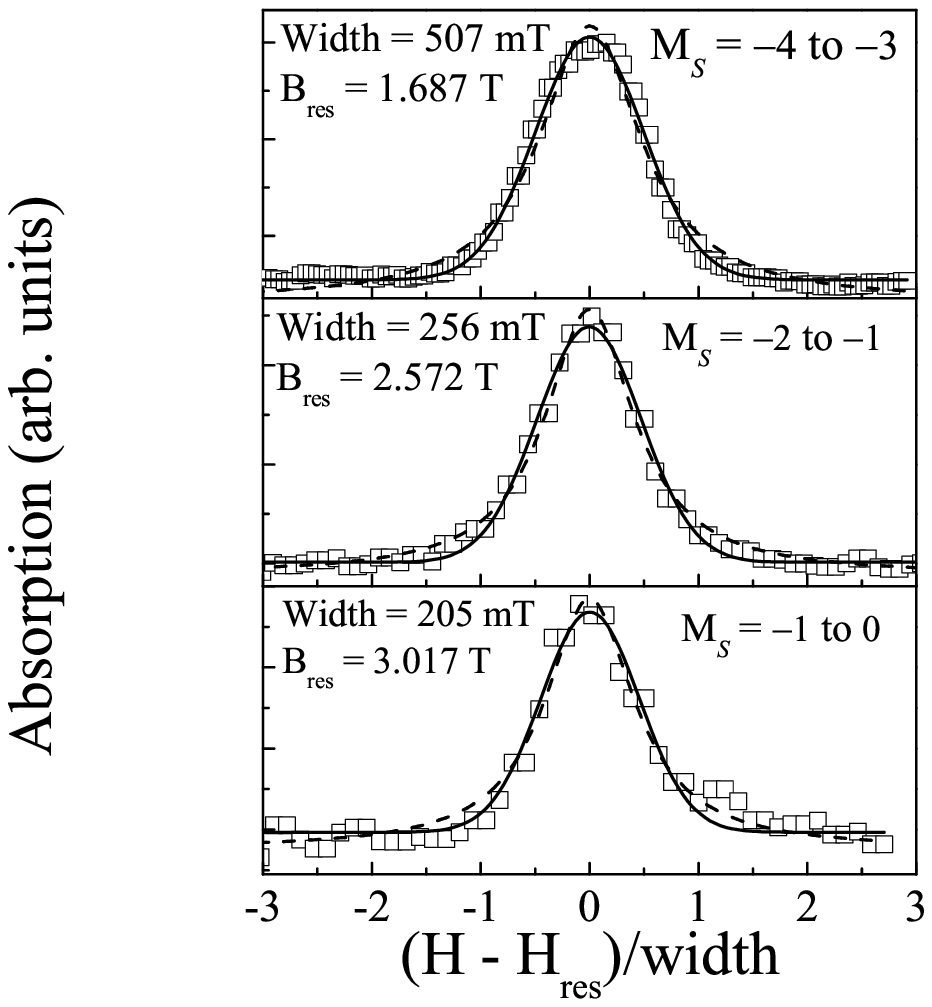,width=120mm}}
\bigskip
\caption{Hill {\em et al.}} \label{Fig. 4}
\end{figure}

\clearpage

\begin{figure}
\centerline{\epsfig{figure=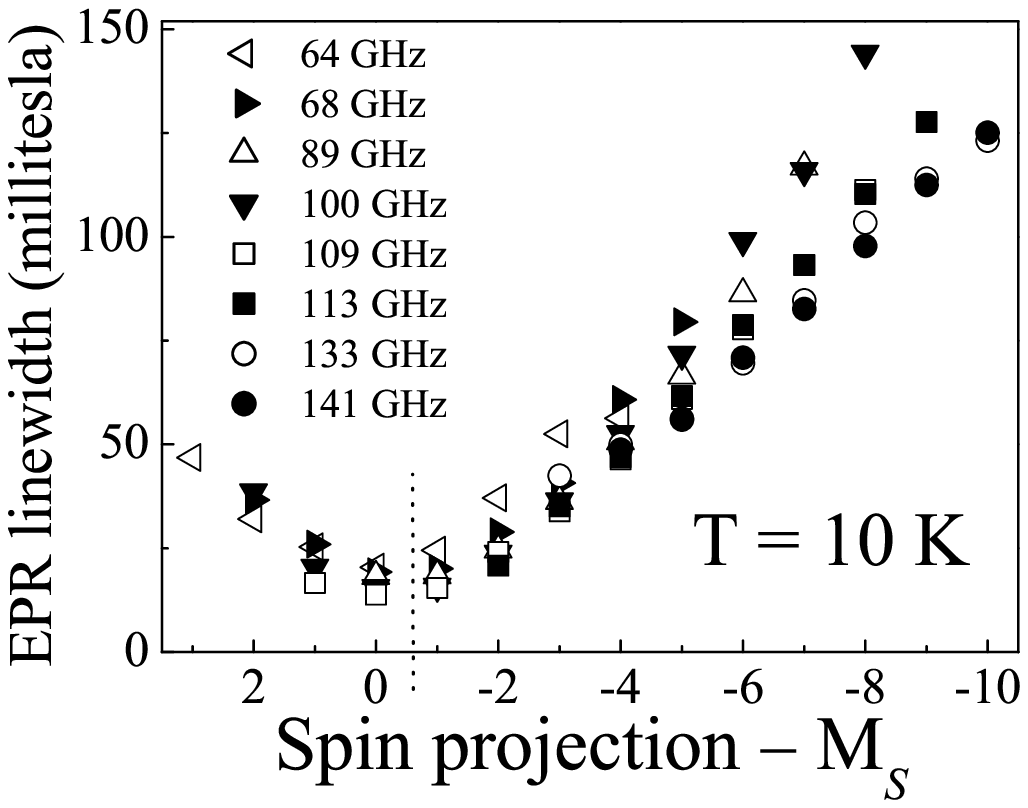,width=160mm}}
\bigskip
\caption{Hill {\em et al.}} \label{Fig. 5}
\end{figure}

\clearpage

\begin{figure}
\centerline{\epsfig{figure=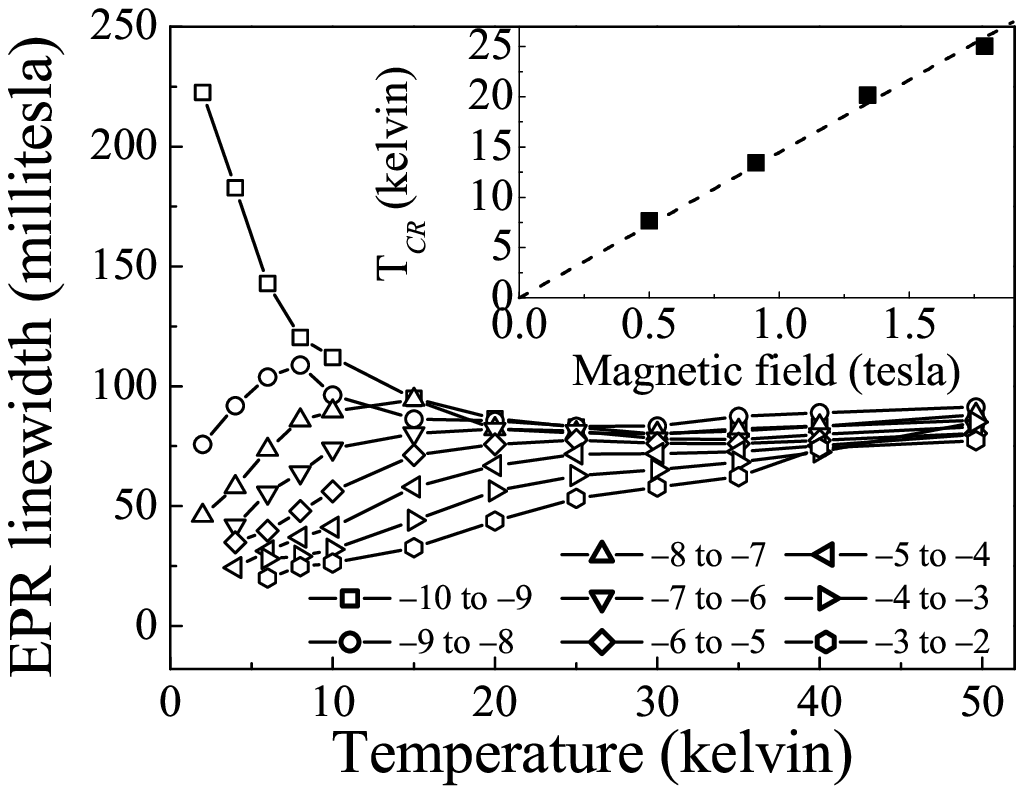,width=160mm}}
\bigskip
\caption{Hill {\em et al.}} \label{Fig. 6}
\end{figure}

\clearpage

\begin{figure}
\centerline{\epsfig{figure=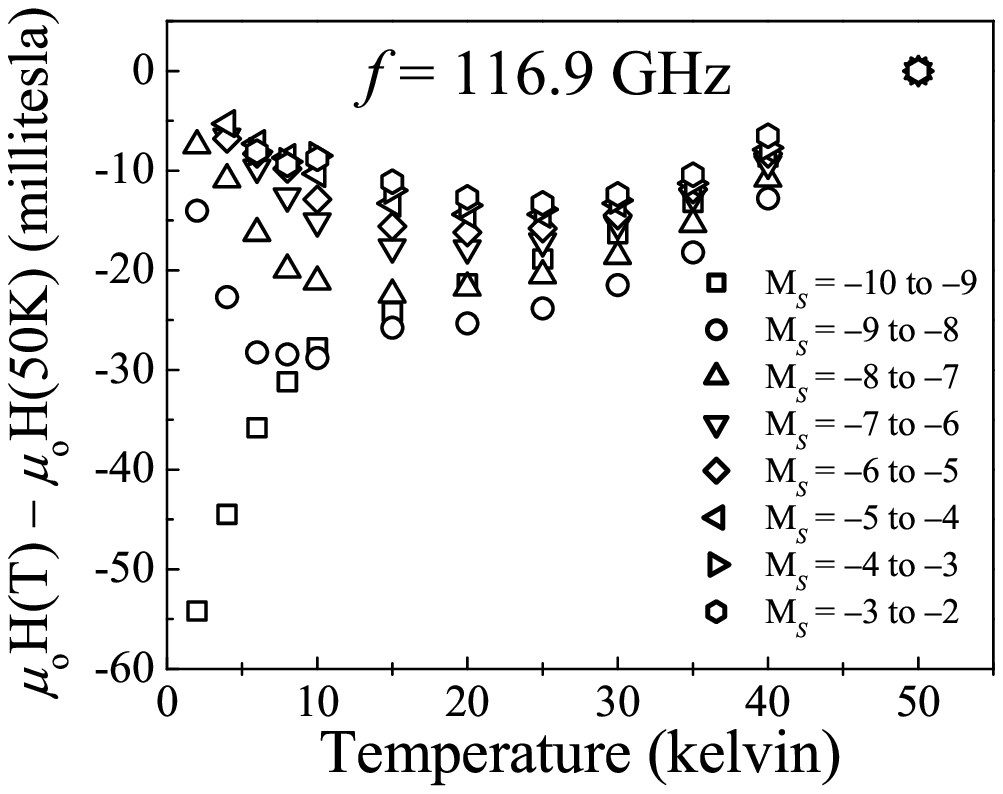,width=160mm}}
\bigskip
\caption{Hill {\em et al.}} \label{Fig. 7}
\end{figure}

\clearpage

\begin{figure}
\centerline{\epsfig{figure=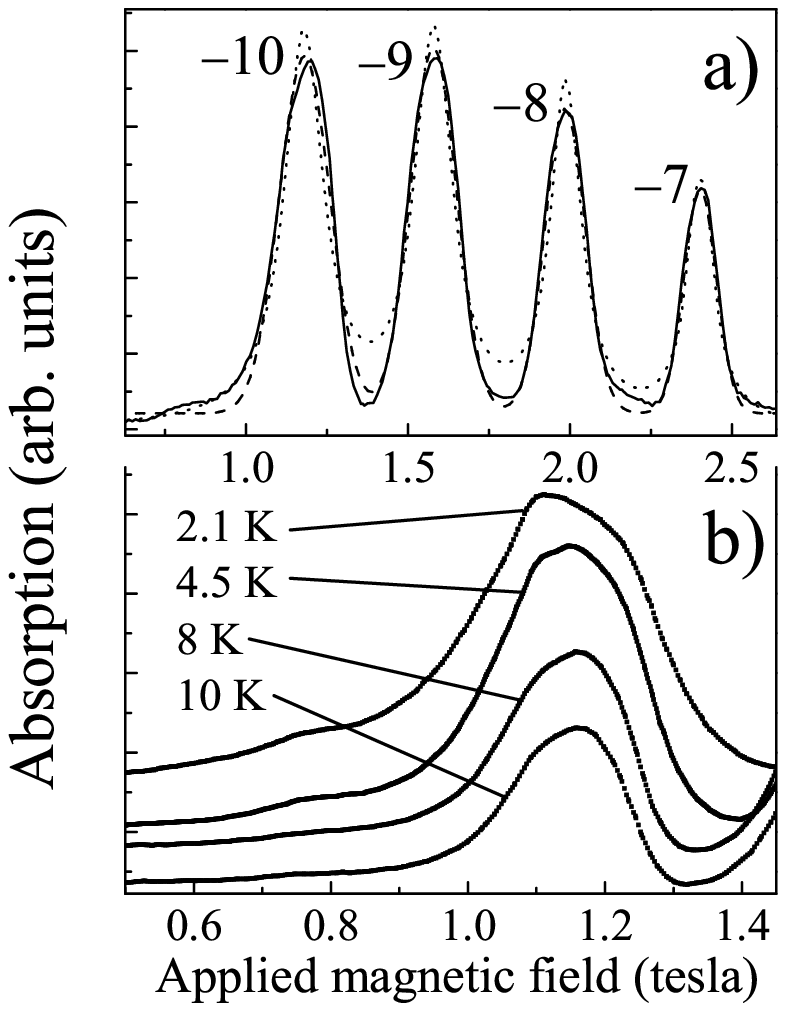,width=160mm}}
\bigskip
\caption{Hill {\em et al.}} \label{Fig. 8}
\end{figure}

\clearpage

\begin{figure}
\centerline{\epsfig{figure=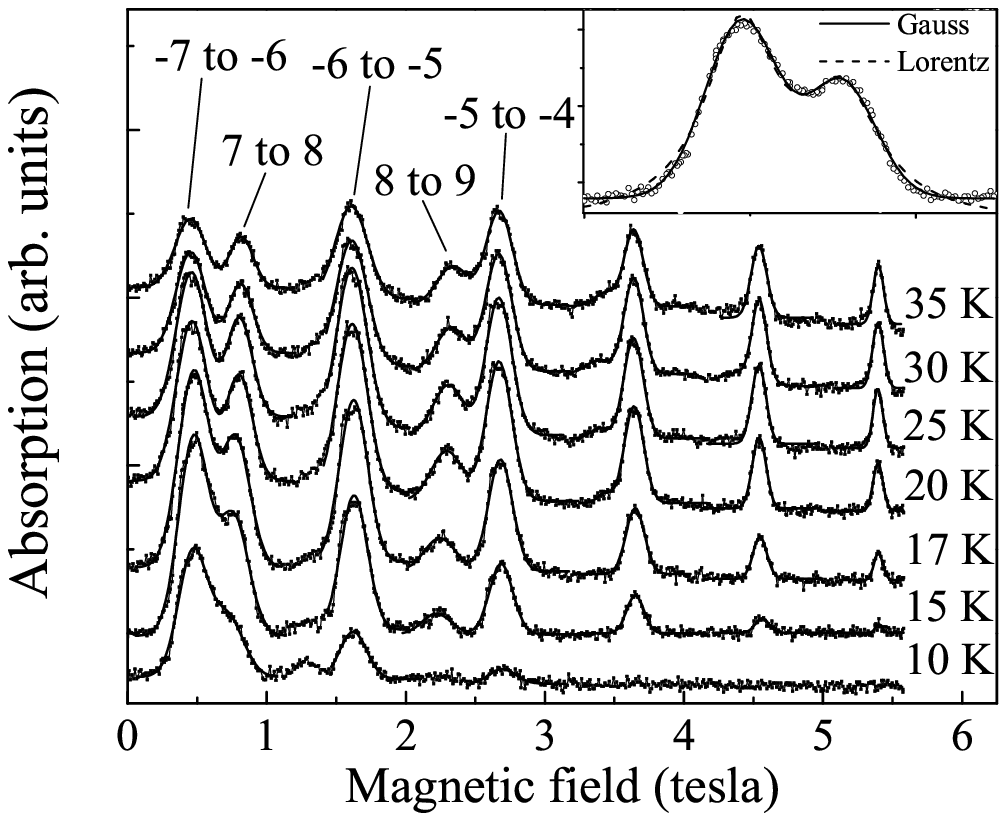,width=160mm}}
\bigskip
\caption{Hill {\em et al.}} \label{Fig. 9}
\end{figure}

\clearpage

\begin{figure}
\centerline{\epsfig{figure=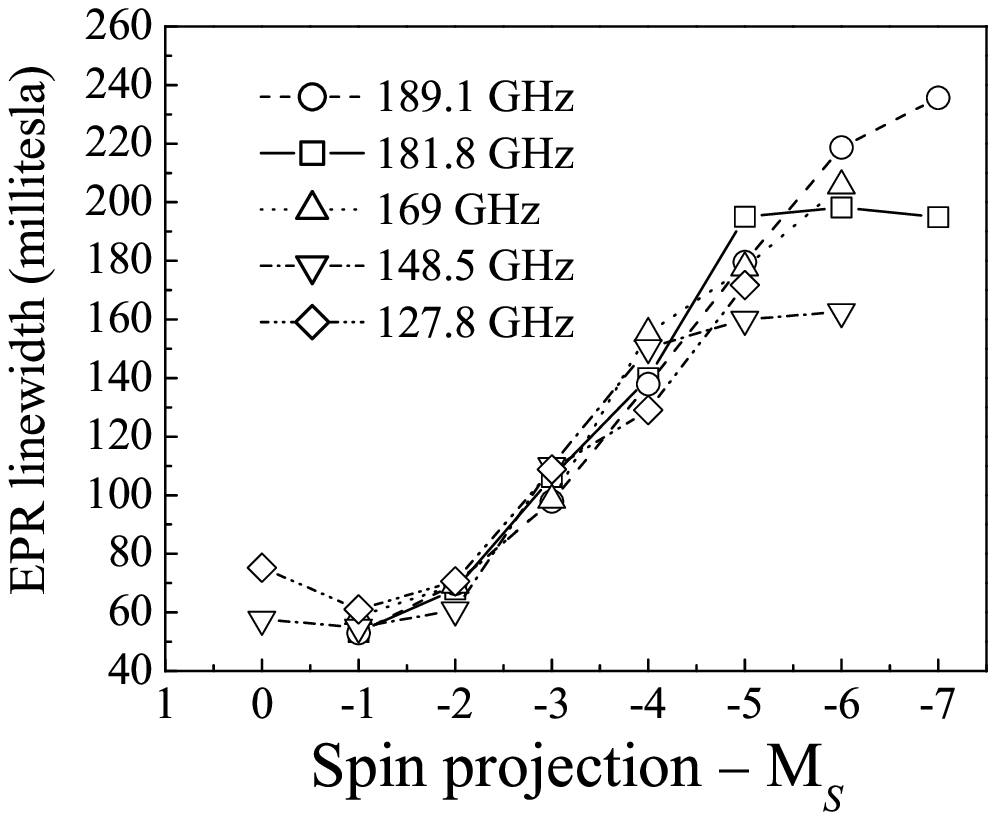,width=160mm}}
\bigskip
\caption{Hill {\em et al.}} \label{Fig. 10}
\end{figure}

\clearpage

\begin{figure}
\centerline{\epsfig{figure=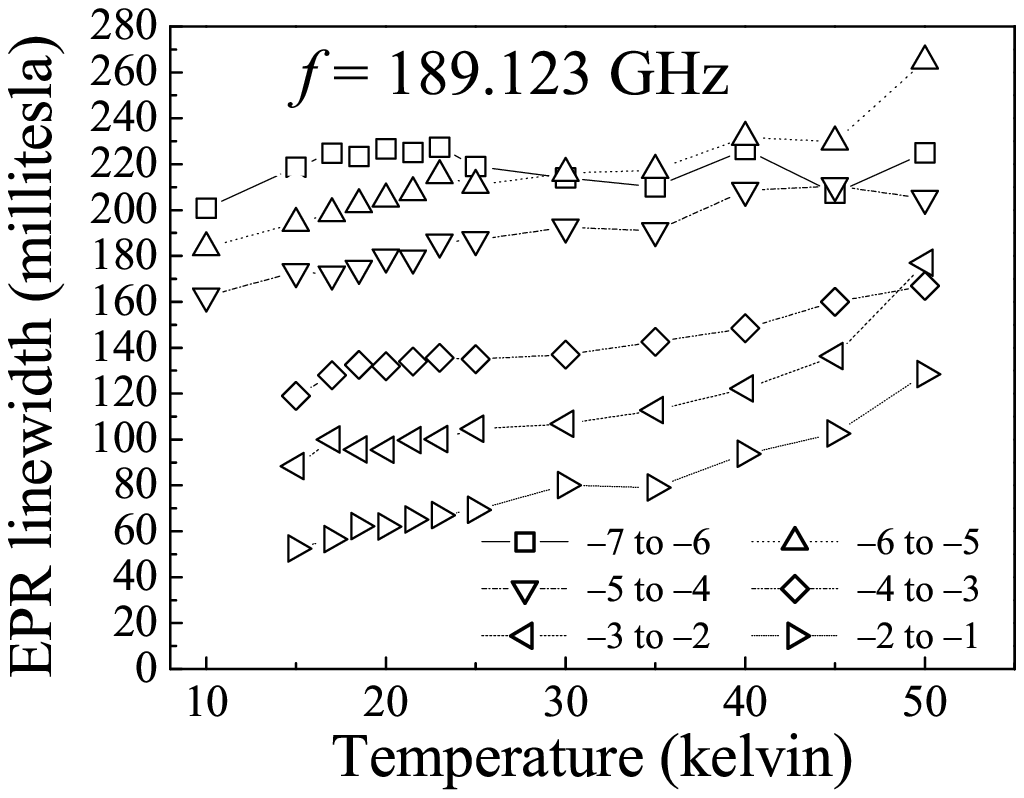,width=160mm}}
\bigskip
\caption{Hill {\em et al.}} \label{Fig. 11}
\end{figure}

\clearpage

\begin{figure}
\centerline{\epsfig{figure=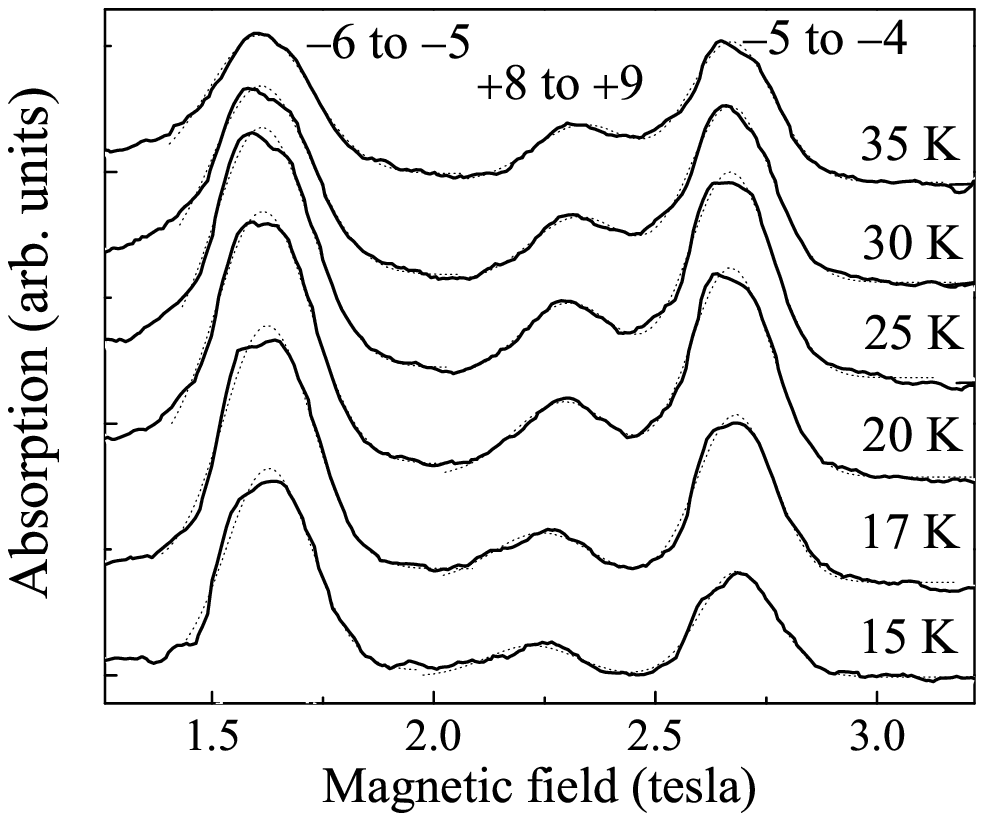,width=160mm}}
\bigskip
\caption{Hill {\em et al.}} \label{Fig. 12}
\end{figure}


\begin{references}
\bibitem{extensions}
This frequency coverage will soon be extended to $>600$ GHz.
\bibitem{simplify}
The dipolar calculation makes the rather bold simplification of
assuming the SMMs to be distributed on a simple cubic lattice.
\bibitem{age}
Very recent measurements on this second sample were made
approximately one year after the original experiments were
performed on the first sample (Figs. 2 thru 7). During this time,
the samples were stored in a refrigerator ($\sim 4^o$C). It is
possible that the samples lost solvent molecules during this
period, as evidenced by the changes in lineshape which are
discussed in main text of this article.
\end{references}
\end{document}